\documentclass{mpe_report}

\usepackage{psfig,graphicx,epsfig}
\usepackage{color}
\usepackage{amsmath,amssymb,epic,eepic,array}
\usepackage{natbib}

\begin{document}

\pagenumbering{arabic}
\setcounter{page}{64}

 \renewcommand{\FirstPageOfPaper }{ 64}\renewcommand{\LastPageOfPaper }{ 67}

\def\aj{AJ}%
\def\actaa{Acta Astron.}%
\def\araa{ARA\&A}%
\def\apj{ApJ}%
\def\apjl{ApJ}%
\def\apjs{ApJS}%
\def\ao{Appl.~Opt.}%
\def\apss{Ap\&SS}%
\def\aap{A\&A}%
\def\aapr{A\&A~Rev.}%
\def\aaps{A\&AS}%
\def\azh{AZh}%
\def\baas{BAAS}%
\def\bac{Bull. astr. Inst. Czechosl.}%
\def\caa{Chinese Astron. Astrophys.}%
\def\cjaa{Chinese J. Astron. Astrophys.}%
\def\icarus{Icarus}%
\def\jcap{J. Cosmology Astropart. Phys.}%
\def\jrasc{JRASC}%
\def\mnras{MNRAS}%
\def\memras{MmRAS}%
\def\na{New A}%
\def\nar{New A Rev.}%
\def\pasa{PASA}%
\def\pra{Phys.~Rev.~A}%
\def\prb{Phys.~Rev.~B}%
\def\prc{Phys.~Rev.~C}%
\def\prd{Phys.~Rev.~D}%
\def\pre{Phys.~Rev.~E}%
\def\prl{Phys.~Rev.~Lett.}%
\def\pasp{PASP}%
\def\pasj{PASJ}%
\def\qjras{QJRAS}%
\def\rmxaa{Rev. Mexicana Astron. Astrofis.}%
\def\skytel{S\&T}%
\def\solphys{Sol.~Phys.}%
\def\sovast{Soviet~Ast.}%
\def\ssr{Space~Sci.~Rev.}%
\def\zap{ZAp}%
\def\nat{Nature}%
\def\iaucirc{IAU~Circ.}%
\def\aplett{Astrophys.~Lett.}%
\def\apspr{Astrophys.~Space~Phys.~Res.}%
\def\bain{Bull.~Astron.~Inst.~Netherlands}%
\def\fcp{Fund.~Cosmic~Phys.}%
\def\gca{Geochim.~Cosmochim.~Acta}%
\def\grl{Geophys.~Res.~Lett.}%
\def\jcp{J.~Chem.~Phys.}%
\def\jgr{J.~Geophys.~Res.}%
\def\jqsrt{J.~Quant.~Spec.~Radiat.~Transf.}%
\def\memsai{Mem.~Soc.~Astron.~Italiana}%
\def\nphysa{Nucl.~Phys.~A}%
\def\physrep{Phys.~Rep.}%
\def\physscr{Phys.~Scr}%
\def\planss{Planet.~Space~Sci.}%
\def\procspie{Proc.~SPIE}%
\let\astap=\aap
\let\apjlett=\apjl
\let\apjsupp=\apjs
\let\applopt=\ao

\title{Comparison of Giant Radio Pulses in Young Pulsars and Millisecond Pulsars}
\author{A. S\l{}owikowska\inst{1, 3} \and A. Jessner\inst{2} \and G. Kanbach\inst{3} \and B. Klein\inst{2}}  
\institute{Copernicus Astronomical Center, Rabia\'nska 8, 87-100 Toru\'n, Poland
\and
Max-Planck-Institut f\"ur Radioastronomie, Auf dem H\"ugel 69, 53121 Bonn, Germany
\and
Max-Planck-Institut f\"ur extraterrestrische Physik, Giessenbachstra{\ss}e 1, 85740 Garching, Germany}
\maketitle

\begin{abstract}
Pulse-to-pulse intensity variations are a common property of pulsar radio emission.
For some of the objects single pulses are often 10-times stronger than their average
pulse. The most dramatic events are so-called giant radio pulses (GRPs). They can be
thousand times stronger than the regular single pulses from the pulsar. Giant pulses are
a rare phenomenon, occurring in very few pulsars which split into two groups. The first
group contains very young and energetic pulsars like the Crab pulsar, and its twin (PSR B0540-69) in the Large Magellanic Cloud (LMC), while the second group is represented by old, recycled millisecond pulsars like PSR B1937+21, PSR B1821-24, PSR B1957+20 and PSR J0218+4232 (the only millisecond pulsar detected in gamma-rays). We compare the characteristics of GRPs for these two pulsar groups. Moreover, our latest findings of new features in the Crab GRPs are presented. Analysis of our Effelsberg data at 8.35~GHz shows that GRPs do occur in all phases of its ordinary radio emission, including the phases of the two high frequency components (HFCs) visible only between 5 and 9~GHz.
\end{abstract}

\section{Introduction}
The `giant pulses' term is reserved for individual radio pulses that are 10--20 or more times stronger than the mean pulse energy. The Crab pulsar was discovered through showing this phenomenon \citep{Staelin1968}. For a long time it was the only pulsar known to emit giant radio pulses (GRPs). Within the group of $\sim 1700$ already known radio pulsars only a few pulsars are found to emit GRPs. These pulsars belong to two different groups of pulsars, some represents the classical pulsars, some of them the millisecond pulsars.

\section{Young pulsars}
\subsection*{The Crab Pulsar}
The occurrence  of sporadic radio emission  of very strong pulses by NP 0532 (the very first name of the Crab  pulsar) has been known since its discovery by \citet*{Staelin1968}. They discovered the dispersed pulse signals from the Crab Nebula. These strong pulses  were  not   periodic,  therefore  the rotational period of  the Crab was not established.  However, considering the source  as a  periodic one, they were  able to  give the upper limit of 0.13~s for its period. A single pulse of average  flux density can not be detected because of the high radio emission from the plerionic nebula. The strongest pulses exceed the total radio emission from the nebula itself by an order of magnitude. The average pulse profile is determined by averaging the data from thousands  of sequentially recorded single pulses. Following studies  of its pulse  shape as a  function of radio  frequency have shown that Crab is  very unique \citep[see e.g.][Figs.~1 and 2; hereafter MH96]{Moffett1996}.  The results show that the averaged radio pulse profile consists of two main components: an intense but narrow `main pulse' (MP), lasting about $250~\rm \mu s$  with respect to the 33~ms pulsar period, and a broader and weaker `interpulse' (IP) following  the MP by  13.37 milliseconds. These two main  components have counterpart non-thermal  emission  from the  infrared  to the  gamma-ray energies. IP does  not occur  exactly between successive  main pulses;  the phase  separation between the MP and the IP  is $\sim 0.4$ and decreases nearly continuously as a function of energy \citep{Eikenberry1997}. At lower radio frequencies (300--600 MHz) another broad and  weak component is visible. It precedes the MP by 1.6 milliseconds and is called  a `precursor' (P). In  the average pulse shape at 1.4~GHz  the precursor vanishes, due  to its steep  spectral index, leaving only MP, IP, and a weak  but distinct low frequency component (LFC). It is $\sim 36\degr$ ahead of the MP, therefore it is not coincident with the position of the precursor  apparent at lower frequency. New and additional profile components were  discovered by MH96. They appear in the profiles obtained for frequencies between 5 and 8~GHz.  These two broad components with nearly flat spectrum are referred to as high frequency component 1 and 2, i.e. HFC1  and HFC2,  respectively. The existence  of the extra  components at high  frequency and  their  strange, frequency-dependent  behaviour is  unlike anything seen  in other pulsars.  It can not  easily be explained  by emission from a  simple dipole field  geometry.

For over 25 years,  only the Crab pulsar was known to  emit GRPs. Its individual     GRPs     might     last     just     a     few     nanoseconds \citep{Hankins2003}.  However, they rank  among the  brightest flashes  in the radio sky  reaching peak flux densities  of up to  1500~Jy even at  high radio frequencies. Already four years after  its discovery it has been reported that GRPs  occur   not  only  at   the  MP,  but   also  at  the  IP   phases \citep{Gower1972}. 

Until 2005 the  GRP phenomenon in the Crab pulsar  had been known to  occur exclusively at the phases  of the MP and the IP \citep{Lundgren1995,Sallmen1999,Cordes2004}. In particular, no GRPs had been detected either in  the LFC or at the phases of  the high radio frequency components: HFC1 and HFC2. The situation changed due to results published by \citet{Jessner2005}. In our observations at 8.35~GHz the IP  and both HFCs are clearly visible, whereas  the LFC can be seen as a slight rise above the noise level separated by about 0.1 in phase from the MP, that is not very intense at this frequency as well. Our observations show that GRPs can be found in all phases of ordinary radio emission including HFC1 and HFC2. For all  giant pulses,  i.e. regardless their  phases, the histogram  of their peak strengths at 8.35 GHz can be  described by a power-law with a slope $ -3.34 \pm  0.19$. This result is  consistent with the results   obtained   by   others  \citep[e.g.][]{Lundgren1995}.   Proportional contribution of  a number  of GRPs  to four different  phase components  is as follow: 80\%, 9\%, 7\% and 4\% for IP, MP, HFC1 and HFC2, respectively. Because of the limited statistics we could make a model $(S/N)^{\alpha}$ fit only for the IP component, where a power-law index $ -3.13 \pm 0.22$ was found. It is consistent with the value of $\sim -2.9$ presented  by \citet{Cordes2004}. The MPs from  the Crab pulsar at 146~MHz are  distributed according to a  power-law with the  exponent of -2.5, whereas the  IPs with -2.8 \citep{Argyle1972}.  For comparison at  800 MHz the distribution  (regardless the  phases  of  GRPs) has  a  slope  of  -3.46 \citep{Lundgren1995}. It  should be  noticed that at  low radio  frequency the main contribution to the number of  GRPs comes from the MP component. So far, there is no evidence that the distributions of numbers of GRPs for the HFC1 and the HFC2 differ from each other. However, their slopes  seem to be steeper than the slope of the same distribution for the IP component. An intriguing feature of GRPs observed with the Effelsberg telescope is that sometimes they occur in a single rotation in more than one component (e.g. at the IP, HFC1, and HFC2 phases in the same rotation). 

The average polarisation  characteristics of about 900 GRPs at 8.35~GHz in some  aspects are similar to previously published high  radio frequencies observations, but in some aspects        we       do        observe        significant       differences . All authors \citep[][hereafter: MH99, K04, S05]{Moffett1999, Karastergiou2004,Slowikowska2005radio} find that the relative offset of the position angle (PA)  between IP and HFCs is on the level of $35\degr-45\degr$. However, the PA for the
IP, and  at the same  time for the  HFCs differ for  all authors. They  are as
follow,  for  the IP:  $30\degr$,  $0\degr$, $-30\degr$,  and for  the  HFCs:
$60\degr-70\degr$, $45\degr$, $5\degr-10\degr$ according to MH99, K04, S05,  respectively. Moreover, there is some discrepancy in the degree of polarisation.
Almost 100\% linear polarisation of all three components has  been derived by K04, whereas for MH99 the IP  is polarised in only 50\% and  HFCs in 80-90\%. In
our work (S05) all  components are polarised at the same level  of 70-80\%. This may
be due to the time varying contribution of the nebula to the rotation  measure  of the  pulsar  (\cite{Rankin1988}:  RM=-43 rad  m$^{-2}$, MH99: RM=-46.9  rad m$^{-2}$, \cite{Weisberg2004}:  RM=-58 rad m$^{-2}$). No abrupt sweeps in PA are found within pulse components. The S/N ratio was too low to derive reliable values of polarisation degree and angle for the LFC and MP components. Some new and additional features of GRPs at MP and IP phases are presented by Eilek \& Hankins (2006, this proceedings). They observed the GRPs with the 2.5~GHz bandwidth (8--10.5~GHz) and found that giant interpulses have totally different intensity profiles and dynamic spectra from giant main pulses.

\subsection*{PSR~B0540-69}
GRPs were also found in the Crab-twin, i.e. PSR B0540-69  - pulsar located in the LMC \citep[$P=50~\rm ms$;][]{Johnston2003}. Pulse profiles and spectra of PSR B0540-69 and PSR B0531+21 differ significantly. The PSR B0540-69 spectrum has very similar shape as the spectral shape of the Class-II MSPs (see Sec. \ref{msps}), and not as the Crab spectrum.  The Crab pulse profile shows a sharp double-peak structure, whereas the profile of LMC pulsar consists of a single broad peak. However, this broad pulse peak might be formed as a superposition of two Gaussian components separated of about 0.2 in phase \citep{dePlaa2003}. The giant pulses occur 6.7~ms before and 5.0~ms after the mid-point of the X-ray profile and they follow the power-law distribution. From performed simultaneous X-ray and radio observations it is known that there is no significant increase in the X-ray pulse profile at the time of GRPs \citep{Johnston2004}. 

\section{Millisecond pulsars (MSPs)}
\label{msps}
Interestingly three millisecond pulsars showing GRPs phenomena form an individual class -
the Class-II of MSPs \citep[after][based on observational X-ray characteristics]{Kuiper2003}.
These pulsars: B1937+12 (P=  $1.56~\rm{ms}$), B1821-24 (P=  $3.05~\rm{ms}$), J0218+4232 (P=  $2.32~\rm{ms}$) have high X-ray luminosities, hard power-law shaped X-ray spectra and their X-ray pulses consist of two narrow components separated in phase by 0.5. All of these pulsars have a very high value of the magnetic field at the light cylinder, $B_{LC}$ \citep{Cognard1996}. Within the group of MSPs, the third highest $B_{LC}$ is found for the millisecond pulsar B1957+20 - being also a potential source of giant pulse emission. It has a high X-ray luminosity, but no X-ray pulsation have been detected from this source so far. A new population of short-duration pulses from B1957+20 is reported by \cite{Knight2006a}. They detected only four of this kind of pulses in 8003~s of observations. They have a sub-microsecond timescale, are several times stronger than the mean pulse, and are coincident with the main emission component. It is worth to mention that this pulsar has a very strong wind that ablates the gas of its companion. Recently, \cite{Knight2005} detected GRPs of up to 64 times the mean pulse energy from PSR BJ1823-3021A ($P= 5.4~\rm ms$). This pulsar is located in the globular cluster NGC 6624 and has the lowest value of the characteristics age of all known MSPs, i.e 26~Myr \citep[period derivative can be perturbed by gravitational interaction in the cluster and consequently can give different value of $\tau_{char} = P/(2\dot{P})$, for the discussion see][]{Knight2005}. As an exception it has not been detected in X-rays. Moreover, nearly all its GRPs are distributed within the trailing half of the main component of the integrated pulse profile. This is in contrast to the location of for example GRPs from PSR B1937+21, that are clustered around the extreme trailing edge of the pulse components. Viewing geometry rather than a different emission mechanism can be responsible for such a difference.  

\subsection*{PSR B1937+12}
The existence of large pulses from PSR B1937+21 was firstly noted by \cite{Wolszczan1984}, but at that time this fact received only a little attention. About ten years later  \cite{Sallmen1995} presented the first analysis of properties of these giant pulses. They noted that GRPs are located on the trailing edges of both main pulse and interpulse. 
This study was later followed by a more extensive one performed by \cite{Cognard1996}.
The GRPs distribution has a power-law shape. These strong pulses are narrower than the averaged one and are systematically delayed by $\sim 40-50~\rm \mu s$. Moreover, many of them are nearly 100\% circularly polarised, while the averaged main pulse is more than 50\% linearly polarised and interpulse at about 13\%; both with little if any circular polarisation.  Observations performed by \cite{Romani2001} provided only 19 bins per pulse profile, however they clearly detected GRPs distributions in both, main and interpulse, components. Again the delay of time of arrival of GRPs relative to the average emission has been confirmed. In both components GRP peaks occur approximately 1 bin after the corresponding peak of the integrated pulse profile. GRPs of PSR B1937+21 are extremely short, with duration below 15~ns, the strongest one has the flux density of 65 000 Jy \citep{Soglasnov2004}. These are the shortest pulses found so far in any pulsar after those of the Crab pulsar \citep{Hankins2003}. 

\subsection*{PSR B1821-24}
This pulsar has a complex radio pulse morphology. It consists of three peaks (P1, P2 and P3) and emission over the whole pulse period. It pulsates in different modes. Its GRPs usually occur in the P1 and P3 phase windows and are concentrated in a narrow phase window coincident with the power-law non-thermal pulse seen in hard X-rays \citep{Romani2001}. In case of the radio pulse profile obtained only from GRPs the P1 emission lags that of the integrated pulse profile by $\sim 80\rm \mu s$. Recently, the polarisation of the GRPs has been reported by  \cite{Knight2006b}. The polarimetric properties of GRPs are completely different to the pulses of ordinary emission.
At $\sim$1.4~GHz the averaged emission of P1 and P2 is  72\% and 96\% linearly polarised, respectively. P3 is not polarised at all. No circular polarisation was found in any component. In contrast, the giant pulses of PSR B1821-24 are highly elliptically polarised, some are even 100\% elliptically polarised. The average polarisation fraction is 79\%.
Their position angles are random and seem to have no preferred orientation. This might be caused by a propagation effect, or it might be intrinsic to the emission mechanism of giant pulses. 

\subsection*{PSR J0218+4232}
For the first time GRPs (only three events) of PSR J0218+4232 were reported by \cite{Joshi2004}, however \cite{Knight2006a} share the opinion that they are rather controversial. The second group of authors describe the analysis of 155 such events. Again, the pulses are very narrow, and are align in phase with the non-thermal X-ray emission, i.e. they occur roughly at the minima of the integrated radio pulse profile. Therefore, they do not contribute to the ordinary radio emission. This strong correlation between phases of GRPs and X-ray maxima confirms that the two emission processes originate in similar regions of the pulsar magnetosphere. The distribution of pulse energies has a power-law shape. However, only 3 over 139 have energies greater than ten times the average. No polarisation information is available. 

\section{Conclusions}
The Class-II MSPs together with the Crab pulsar and its twin from the LMC are among the top seven pulsars with the highest magnetic field strengths near the light cylinder \citep[the sixth one is PSR B1957+20, and the seventh is the radio-quiet one - PSR J0537-6910;][]{Cognard1996, Kramer2004}. However, new observations do not support very strongly the hypothesis that strong magnetic field at the light cylinder is a critical condition for giant pulse activity.

The Crab pulsar results obtained so far by  us suggest that physical conditions  in the regions responsible  for HFCs emission might  be similar to those  in  the main
pulse and interpulse emission regions. This idea is supported not only  by our
detection  of GRPs  phenomenon  at LFC  and  HFCs phases,  but  also by  their
polarisation characteristics.  Still, the origin of  HFCs remains an open question. One of the propositions is inward emission from outer gaps  which may produce two additional peaks at requested phases in the light curve of the Crab pulsar \citep{Cheng2000}.

Flux density distribution of GRPs that have been detected from different  pulsars generally  follows power-law  statistics.  GRPs  are  believed   (to some extend) to  be associated  with non-thermal  high  energy emission  (e.g. PSR  B0531+21,  PSR B1937+21,  PSR B1821-24,  and  PSR B0540-69).  The  Crab pulsar  was  the  best example of  this association for a  long time. From the  group of  millisecond pulsars a strong confirmation of this  idea comes from the observations  of PSR B1821-24.  \citet{Romani2001} as well as \citet{Knight2006b} found  that GRPs of this  pulsar are concentrated in a  narrow phase window coincident with the non-thermal pulses seen in  hard X-rays, but occur on the trailing edges  of the radio components.  Recently,  more and more features of pulsar radiation showing correlations between radio and high energy emission are observed.  For 
example phase  alignment between some  optical polarisation  features  and  radio intensity profile \citep{SlowikowskaPhD}. Furthermore, it was found that there is a
correlation  between  X-ray and  radio  pulses  for Vela \citep{Lommen2006}, whereas  \citet{Shearer2003}  have  detected  a  correlation  between optical emission and  GRPs emission in the Crab pulsar.  They found that  optical pulses coincident with GRPs were of  about  3\% brighter  on average.   On the other hand, the characteristics of some pulsars \citep[PSR B1133+16;][]{Kramer2003} and the  discovery of the  Crab's giant radio  pulses at phases where   no high energy emission is known, do not match to this picture.

\begin{acknowledgements}
AS acknowledges the support from the Polish grant 2P03D.004.24. AS also acknowledges support from the Deutscher Akademischer Austausch Dienst (DAAD).
\end{acknowledgements}
  
\bibliographystyle{aa}
\bibliography{as_grp.bib}

\begin{thebibliography}{34}
\expandafter\ifx\csname natexlab\endcsname\relax\def\natexlab#1{#1}\fi

\bibitem[{{Argyle} \& {Gower}(1972)}]{Argyle1972}
{Argyle}, E. \& {Gower}, J.~F.~R. 1972, in Bulletin of the American
  Astronomical Society, 216

\bibitem[{{Cheng} {et~al.}(2000){Cheng}, {Ruderman}, \& {Zhang}}]{Cheng2000}
{Cheng}, K.~S., {Ruderman}, M., \& {Zhang}, L. 2000, \apj, 537, 964

\bibitem[{{Cognard} {et~al.}(1996){Cognard}, {Shrauner}, {Taylor}, \&
  {Thorsett}}]{Cognard1996}
{Cognard}, I., {Shrauner}, J.~A., {Taylor}, J.~H., \& {Thorsett}, S.~E. 1996,
  \apjl, 457, L81+

\bibitem[{{Cordes} {et~al.}(2004){Cordes}, {Bhat}, {Hankins}, {McLaughlin}, \&
  {Kern}}]{Cordes2004}
{Cordes}, J.~M., {Bhat}, N.~D.~R., {Hankins}, T.~H., {McLaughlin}, M.~A., \&
  {Kern}, J. 2004, \apj, 612, 375

\bibitem[{{de Plaa} {et~al.}(2003){de Plaa}, {Kuiper}, \&
  {Hermsen}}]{dePlaa2003}
{de Plaa}, J., {Kuiper}, L., \& {Hermsen}, W. 2003, \aap, 400, 1013

\bibitem[{{Eikenberry} \& {Fazio}(1997)}]{Eikenberry1997}
{Eikenberry}, S.~S. \& {Fazio}, G.~G. 1997, \apj, 476, 281

\bibitem[{{Gower} \& {Argyle}(1972)}]{Gower1972}
{Gower}, J.~F.~R. \& {Argyle}, E. 1972, \apjl, 171, L23

\bibitem[{{Hankins} {et~al.}(2003){Hankins}, {Kern}, {Weatherall}, \&
  {Eilek}}]{Hankins2003}
{Hankins}, T.~H., {Kern}, J.~S., {Weatherall}, J.~C., \& {Eilek}, J.~A. 2003,
  \nat, 422, 141

\bibitem[{{Jessner} {et~al.}(2005){Jessner}, {S{\l}owikowska}, {Klein},
  {Lesch}, {Jaroschek}, {Kanbach}, \& {Hankins}}]{Jessner2005}
{Jessner}, A., {S{\l}owikowska}, A., {Klein}, B., {et~al.} 2005, Advances in
  Space Research, 35, 1166

\bibitem[{{Johnston} \& {Romani}(2003)}]{Johnston2003}
{Johnston}, S. \& {Romani}, R.~W. 2003, \apjl, 590, L95

\bibitem[{{Johnston} {et~al.}(2004){Johnston}, {Romani}, {Marshall}, \&
  {Zhang}}]{Johnston2004}
{Johnston}, S., {Romani}, R.~W., {Marshall}, F.~E., \& {Zhang}, W. 2004,
  \mnras, 355, 31

\bibitem[{{Joshi} {et~al.}(2004){Joshi}, {Kramer}, {Lyne}, {McLaughlin}, \&
  {Stairs}}]{Joshi2004}
{Joshi}, B.~C., {Kramer}, M., {Lyne}, A.~G., {McLaughlin}, M.~A., \& {Stairs},
  I.~H. 2004, in IAU Symp. 218, ed. F.~{Camilo} \& B.~M. {Gaensler}, 319

\bibitem[{{Karastergiou} {et~al.}(2004){Karastergiou}, {Jessner}, \&
  {Wielebinski}}]{Karastergiou2004}
{Karastergiou}, A., {Jessner}, A., \& {Wielebinski}, R. 2004, in IAU Symp. 218,
  ed. F.~{Camilo} \& B.~M. {Gaensler}, 329--330

\bibitem[{{Knight} {et~al.}(2005){Knight}, {Bailes}, {Manchester}, \&
  {Ord}}]{Knight2005}
{Knight}, H.~S., {Bailes}, M., {Manchester}, R.~N., \& {Ord}, S.~M. 2005, \apj,
  625, 951

\bibitem[{{Knight} {et~al.}(2006{\natexlab{a}}){Knight}, {Bailes},
  {Manchester}, \& {Ord}}]{Knight2006b}
{Knight}, H.~S., {Bailes}, M., {Manchester}, R.~N., \& {Ord}, S.~M.
  2006{\natexlab{a}}, astro-ph/0608155

\bibitem[{{Knight} {et~al.}(2006{\natexlab{b}}){Knight}, {Bailes},
  {Manchester}, {Ord}, \& {Jacoby}}]{Knight2006a}
{Knight}, H.~S., {Bailes}, M., {Manchester}, R.~N., {Ord}, S.~M., \& {Jacoby},
  B.~A. 2006{\natexlab{b}}, \apj, 640, 941

\bibitem[{{Kramer}(2004)}]{Kramer2004}
{Kramer}, M. 2004, in IAU Symposium, ed. F.~{Camilo} \& B.~M. {Gaensler}

\bibitem[{{Kramer} {et~al.}(2003){Kramer}, {Karastergiou}, {Gupta}, {Johnston},
  {Bhat}, \& {Lyne}}]{Kramer2003}
{Kramer}, M., {Karastergiou}, A., {Gupta}, Y., {et~al.} 2003, \aap, 407, 655

\bibitem[{Kuiper \& Hermsen(2003)}]{Kuiper2003}
Kuiper, L. \& Hermsen, W. 2003, astro-ph/0312204

\bibitem[{Lommen {et~al.}(2006)Lommen, Donovan, Gwinn, Arzoumanian, Harding,
  Strickman, Dodson, McCulloch, \& Moffett}]{Lommen2006}
Lommen, A., Donovan, J., Gwinn, C., {et~al.} 2006, astro-ph/0611450

\bibitem[{{Lundgren} {et~al.}(1995){Lundgren}, {Cordes}, {Ulmer}, {Matz},
  {Lomatch}, {Foster}, \& {Hankins}}]{Lundgren1995}
{Lundgren}, S.~C., {Cordes}, J.~M., {Ulmer}, M., {et~al.} 1995, \apj, 453, 433

\bibitem[{{Moffett} \& {Hankins}(1996)}]{Moffett1996}
{Moffett}, D.~A. \& {Hankins}, T.~H. 1996, \apj, 468, 779

\bibitem[{{Moffett} \& {Hankins}(1999)}]{Moffett1999}
{Moffett}, D.~A. \& {Hankins}, T.~H. 1999, \apj, 522, 1046

\bibitem[{{Rankin} {et~al.}(1988){Rankin}, {Campbell}, {Isaacman}, \&
  {Payne}}]{Rankin1988}
{Rankin}, J.~M., {Campbell}, D.~B., {Isaacman}, R.~B., \& {Payne}, R.~R. 1988,
  \aap, 202, 166

\bibitem[{{Romani} \& {Johnston}(2001)}]{Romani2001}
{Romani}, R.~W. \& {Johnston}, S. 2001, \apjl, 557, L93

\bibitem[{{Sallmen} \& {Backer}(1995)}]{Sallmen1995}
{Sallmen}, S. \& {Backer}, D.~C. 1995, in ASP Conf. Ser. 72, ed. A.~S.
  {Fruchter}, M.~{Tavani}, \& D.~C. {Backer}, 340

\bibitem[{{Sallmen} {et~al.}(1999){Sallmen}, {Backer}, {Hankins}, {Moffett}, \&
  {Lundgren}}]{Sallmen1999}
{Sallmen}, S., {Backer}, D.~C., {Hankins}, T.~H., {Moffett}, D., \& {Lundgren},
  S. 1999, \apj, 517, 460

\bibitem[{{Shearer} {et~al.}(2003){Shearer}, {Stappers}, {O'Connor}, {Golden},
  {Strom}, {Redfern}, \& {Ryan}}]{Shearer2003}
{Shearer}, A., {Stappers}, B., {O'Connor}, P., {et~al.} 2003, Science, 301, 493

\bibitem[{S\l{}owikowska(2006)}]{SlowikowskaPhD}
S\l{}owikowska, A. 2006, PhD thesis, Nicolaus Copernicus Astronomical Center,
  Warsaw

\bibitem[{{S{\l}owikowska} {et~al.}(2005){S{\l}owikowska}, {Jessner}, {Klein},
  \& {Kanbach}}]{Slowikowska2005radio}
{S{\l}owikowska}, A., {Jessner}, A., {Klein}, B., \& {Kanbach}, G. 2005, in AIP
  Conf. Proc. 801, ed. T.~{Bulik}, B.~{Rudak}, \& G.~{Madejski}, 324--329

\bibitem[{{Soglasnov} {et~al.}(2004){Soglasnov}, {Popov}, {Bartel}, {Cannon},
  {Novikov}, {Kondratiev}, \& {Altunin}}]{Soglasnov2004}
{Soglasnov}, V.~A., {Popov}, M.~V., {Bartel}, N., {et~al.} 2004, \apj, 616, 439

\bibitem[{{Staelin} \& {Reifenstein}(1968)}]{Staelin1968}
{Staelin}, D.~H. \& {Reifenstein}, E.~C. 1968, Science, 162, 1481

\bibitem[{{Weisberg} {et~al.}(2004){Weisberg}, {Cordes}, {Kuan}, {Devine},
  {Green}, \& {Backer}}]{Weisberg2004}
{Weisberg}, J.~M., {Cordes}, J.~M., {Kuan}, B., {et~al.} 2004, \apjs, 150, 317

\bibitem[{{Wolszczan} {et~al.}(1984){Wolszczan}, {Cordes}, \&
  {Stinebring}}]{Wolszczan1984}
{Wolszczan}, A., {Cordes}, J., \& {Stinebring}, D. 1984, in Birth and Evolution
  of Neutron Stars: Issues Raised by Millisecond Pulsars, ed. S.~P. {Reynolds}
  \& D.~R. {Stinebring}

\end{thebibliography}


  \clearpage

\end{document}